\documentclass[reprint,NumberedRefs]{JASA}
\usepackage{amsmath,amssymb,amsfonts}
\usepackage[T1]{fontenc} 
\usepackage{graphicx}
\usepackage{hyperref}

\usepackage{enumitem}
\usepackage{url}
\usepackage{colortbl}

\usepackage{ulem} 

\definecolor{gree}{rgb}{0.0,0.7,0}
\definecolor{brau}{rgb}{0.4,0.5,0}
\definecolor{darkgreen}{rgb}{0,0.6,0}

\newcommand{\rev}[1]{{#1}} 

\begin{document}

\title[]{Digital twins enable full-reference quality assessment of photoacoustic image reconstructions}


\author{Janek Gr\"ohl}
\email{j.groehl@eni-g.de}

\affiliation{Department of Physics, University of Cambridge, Cambridge, U.K.}
\affiliation{Cancer Research UK Cambridge Institute, University of Cambridge, Cambridge, U.K.}
\affiliation{ENI-G, a Joint Initiative of the University Medical Center Göttingen and the Max Planck Institute for Multidisciplinary Sciences, Göttingen, Germany}

\author{Leonid Kunyansky}
\affiliation{Department of Mathematics, University of Arizona, Tucson, USA}

\author{Jenni Poimala}
\affiliation{Department of Technical Physics, University of Eastern Finland, Kuopio, Finland}

\author{Thomas R. Else}
\affiliation{Department of Bioengineering, Imperial College London, London, U.K.}

\author{Francesca Di Cecio}
\affiliation{Department of Physics, University of Cambridge, Cambridge, U.K.}
\affiliation{Cancer Research UK Cambridge Institute, University of Cambridge, Cambridge, U.K.}

\author{\\Sarah E. Bohndiek}
\affiliation{Department of Physics, University of Cambridge, Cambridge, U.K.}
\affiliation{Cancer Research UK Cambridge Institute, University of Cambridge, Cambridge, U.K.}

\author{Ben T. Cox}
\affiliation{Department of Medical Physics and Biomedical Engineering, University College London, London, U.K.}

\author{Andreas Hauptmann}
\email{Andreas.Hauptmann@oulu.fi}
\affiliation{Research Unit of Mathematical Sciences, University of Oulu, Oulu, Finland}
\affiliation{Department of Computer Science, University College London, London, U.K.}








\begin{abstract}
Quantitative comparison of the quality of photoacoustic image reconstruction algorithms remains a major challenge. No-reference image quality measures are often inadequate, but full-reference measures require access to an ideal reference image. While the ground truth is known in simulations, it is unknown {\textit{in vivo}, or} in phantom studies, as the reference depends on both the phantom properties and the imaging system. We tackle this problem by using numerical digital twins of tissue-mimicking phantoms and the imaging system \rev{to perform a quantitative calibration to reduce the \textit{simulation gap}}. {The contributions of this paper are two-fold: First, we use this digital-twin framework to compare multiple state-of-the-art reconstruction algorithms. Second, among these is a Fourier transform-based reconstruction algorithm for circular detection geometries, which we test on experimental data for the first time. Our results demonstrate the usefulness of digital phantom twins by enabling assessment of the accuracy of the numerical forward model and enabling comparison of image reconstruction schemes with full-reference image quality assessment. We show that the Fourier transform-based algorithm yields results comparable to those of iterative time reversal, but at a lower computational cost. \rev{All data and code are publicly available on Zenodo: \url{https://doi.org/10.5281/zenodo.15388429}.}}
\end{abstract}


\maketitle












\section{Introduction}
\label{sec:introduction}

Photoacoustic (PA) imaging (PAI) is a medical imaging modality that promises advances in multiple clinical applications, such as diagnosis and staging of Crohn's disease~\cite{knieling2017multispectral}, Neuromuscular Degenerative Disease~\cite{regensburger2019detection}, Breast Cancer~\cite{manohar2019current}, and multiple others~\cite{assi2023review,lin2022emerging}. The potential of PAI in this context comes from its ability to resolve molecular imaging contrast based on the optical absorption coefficient of the tissue, while simultaneously enabling multi-scale imaging at depths of up to several centimetres in tissue~\cite{beard2011biomedical}. 

From an engineering standpoint, PAI is a coupled-physics modality, combining the high resolution of ultrasound techniques with the high sensitivity of electromagnetic waves to the optical properties of biological tissues.  The region of interest (e.g., a woman's breast in mammography) is irradiated with a short laser pulse. \rev{Light} is partially absorbed by tissues, which raises the temperature of the medium. By a process referred to as thermoelastic expansion, an acoustic wave emerges that is measured by transducers on the object boundary. The propagation of the pressure wave $p(t,x)$ can be modelled by the wave equation 
\begin{equation} \label{E:3Dwave}
\begin{cases}
p_{tt}=c^{2}\Delta p,\quad t\geq 0,\quad x\in \mathbb{R}^{D} \\ 
p(0,x)=p_0(x),\quad p_{t}(0,x)=0,\\
g(t,y_j)=p(t,y_j), \quad j=1,...,N_d%
\end{cases}%
\end{equation}%
where $p_0(x)$ is the initial pressure in the tissues, $c$ is the speed of sound, $y_j$ are the locations of $N_d$ point-like transducers, $g(t,y_j)$ are the measurements made by these transducers, and $D$ is the dimension of the space (in this case $D=3$). This simplified model assumes that \rev{light} absorption happens instantaneously and that the acoustic wave propagates in the open space, i.e., without reflecting from transducers or other parts of the acquisition scheme. It also neglects the absorption and dispersion of acoustic waves in tissues and the frequency and directional responses of real transducers. The \textit{acoustic inverse problem} of PAI consists of reconstructing the initial pressure $p_0$ from the measurements $g(t)$  at the tissue boundary. The \textit{optical inverse problem} then describes reconstructing the optical absorption coefficient $\mu_a$ from $p_0$. Accurate acoustic inversion is thus a fundamental step towards reproducible and quantitative PA data analysis. It is one of the most researched topics in the field of PAI~\cite{poudel2019survey} and many methods to solve the acoustic inverse problem have been proposed, see for example~\cite{xu2005universal, perrot2021so, xu2002pulsed, tarvainen2024quantitative, treeby2010photoacoustic, arridge2016adjoint}.

From a mathematical standpoint, in order to reconstruct $p_0(x)$, which is a function of a 3D variable, measurements of pressure $p(t,x)$ should be performed on a 2D surface, at least partially surrounding the source of acoustic wave~\cite{KuKu2015}. Most PAI image reconstruction algorithms presented in literature are based on this assumption. However, in modern biomedical practice most often one has to deal with the situation where the data are collected by a \rev{sparse arrangement} of detectors and the illuminated region is \rev{only} a part of the whole object. Such is the case with our experimental setup which uses the MSOT inVision-256TF pre-clinical PAI system (iThera Medical GmbH, Munich, Germany) that has an angular coverage of only 270$^\circ$ (see the detailed description in Section~\ref{sec:msot}). Accurate image reconstruction of a 3D image from 2D data is not possible; thus all existing algorithms can produce only approximations of $p_0(x)$. And while all exact reconstructions are alike, each approximate reconstruction is approximate in its own way, thus introducing unique reconstruction artefacts~\cite{rietberg2025artifacts}. One of the goals of this paper is to compare such approximations using both simulated and real data.

One particularly promising acoustic inversion algorithm is an FFT-based reconstruction algorithm that is optimised specifically for circular detection geometries and was first presented by Kunyansky in 2012~\cite{Kun-circ}. {(Note this is different from the well-established FFT reconstruction scheme for planar geometries~\cite{jaeger2007fourier}.)} All computationally non-trivial steps of this algorithm are done using 1D or 2D FFTs, which makes this algorithm very fast. Furthermore, it is straightforward to implement within frameworks like PyTorch and so available to use in learned reconstruction frameworks, such as e.g.~\cite{hauptmann2018model,boink2019partially,hsu2023fast}. The objective of the work presented in this paper is to investigate how the algorithm compares to other state-of-the-art approaches, particularly time reversal~\cite{treeby2010photoacoustic} or model-based image reconstruction~\cite{rosenthal2010fast}. We present a concise summary of the algorithm as part of this study and make it available open source. 
{Additionally, for the first time the FFT-based algorithm for circular geometries is tested on experimental data. This is particularly relevant, since the model assumptions of ideal 2D wave propagation are violated in the experimental system.}

The quantitative evaluation and comparison with other reconstruction algorithms is challenging on experimentally acquired data, however, as the initial pressure distribution that represents the ground truth is usually unknown~\cite{grohl2021deep}. This restricts researchers to simulation studies or no-reference image quality assessment (IQA) measures. While common no-reference IQA measures, such as the full-width at half maximum (FWHM) or the signal-to-noise-ratio (SNR), are well-established to estimate general-purpose image quality {indicators} in natural images, they only have limited applicability to medical images~\cite{breger2024study} and are thus not necessarily indicative of the accuracy of the acoustic inversion.

\begin{figure}[h!tb]
    \centering
    \includegraphics[width=\linewidth]{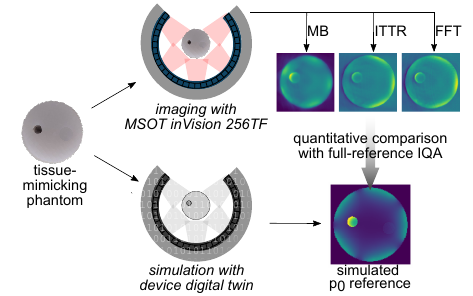}
    \caption{Overview of the proposed evaluation strategy for full-reference IQA of reconstruction algorithms for photoacoustic images. A tissue-mimicking phantom is measured by the imaging device and a reference $p_0(x)$ is simulated using digital twins of the phantom and the device. Reconstructions of different algorithms can then quantitatively be compared against the simulated reference.}
    \label{fig:method}
\end{figure}

{To use the more informative full-reference IQA measures, a reference image of $p_0$ is required to provide a reliable performance measure for the reconstruction quality. In pure simulation studies, $p_0$ is known and can be used as a ground truth, but in experimental settings, this is typically infeasible. State-of-the-art approaches normally use acoustic simulations of binary acoustic pressure maps for the comparison of reconstruction algorithms~\cite{warbal2022performance}. For experimental settings, point sources are often considered~\cite{paltauf2007experimental}, which can be evaluated without knowledge of $p_0$ by characterising the resolution~\cite{van2016comparison}. In almost all cases, however, the underlying optical properties are unknown and the light fluence is only approximated.

This is even true in phantom studies - however carefully the phantom is fabricated and characterised - $p_0$ cannot be known directly as it depends not just on the phantom itself but on the illumination pattern used by the imaging system. If we had a perfect set of measurement data (i.e. complete data) of the phantom, then an exact algorithm could be used to recover the ground truth $p_0$ as a reference to compare against. However, we do not have that data in this case and in most other measurement scenarios. Therefore, to know what the ground truth $p_0$ is, it is necessary to model the phantom and the imaging system (cf. FIG.~\ref{fig:method}).}

{As such, there is an unmet need to test both the applicability of full-reference IQA algorithms for photoacoustic imaging, as well as the comparison of image reconstruction algorithm performance with incomplete data in experimental settings. In this work, we propose a framework based on digital twins of test objects to tackle these problems. We simulate numerical representations of well-characterised imaging phantoms. This allows the comparison of reconstruction results and numerical simulations of $p_0$ \rev{as a means to quantify the remaining \textit{simulation gap} after calibration}. Because of the availability of paired data in this digital twin framework, we can directly compare simulated acoustic pressure measurement data with actual acoustic pressure measurements \rev{and perform a quantitative calibration of the simulation pipeline. The quantitative measures can then} give an essential indication of how accurate the simulated $p_0$ is as the reference - a step that is missing in the experimental comparison of reconstruction algorithms across the state of the art~\cite{choi2020practical}. \rev{This allows for the derivation of a quantifiable measure to judge the reliability of} full-reference IQA measures to determine the reconstruction quality of the algorithms.}

We use these digital twins as an evaluation framework to compare the performance of the FFT-based reconstruction algorithm to other state-of-the-art image reconstruction schemes (FIG.~\ref{fig:method}) and show that its performance is on par while offering computational advantages.

\section{Methods}
\label{sec:methods}

\subsection{Data}

\subsubsection{Phantom Data} We used N=30 cylindrical phantoms with a diameter of 27.5mm, a height of 65-80mm, and an approximate volume of 40-50mL. They were fabricated based on a previously published protocol~\cite{hacker2021copolymer} and are a subset of the phantoms used in a prior publicaton~\cite{grohl2023moving}. Each phantom is piecewise constant and consists of a background cylinder into which imaging targets are added. Rectangular samples with a length of 5.9\,cm, a width of 1.8\,cm, and a thickness ranging between 2mm and 4mm were prepared from each material for optical characterisation. Sample thicknesses were determined at five distinct locations using digital vernier callipers. \\

\subsubsection{Photoacoustic Imaging}
\label{sec:msot}
All data were acquired using the MSOT inVision-256TF pre-clinical PAI system (iThera Medical GmbH, Munich, Germany). It has 256 transducer elements with a 5-MHz centre frequency and 60\% bandwidth, arranged in a circular array of radius 40.5\,mm and angular coverage of 270$^\circ$. More details on the measurement setup can be found in a prior publication~\cite{joseph2017evaluation}. We reconstructed the images using various reconstruction schemes, with their details outlined later. The images were reconstructed into a $300\times300$ pixel grid with a field of view of $32\times32$\,mm.\\

\subsubsection{Digital Twin Simulations} 
\label{sec:digital_twins}

Our phantoms were created in such a way that they had a piecewise-constant material distribution. This allowed us to take samples of each material and characterise them with an in-house double integrating sphere (DIS) system~\cite{hacker2021copolymer} based on the system developed by Pickering et al.~\cite{pickering1993double} and determine the optical parameters, absorption $\mu_a$ and reduced scattering $\mu_s'$, in a wavelength range of 600 to 950\,nm.

{We implemented numerical phantoms by assigning the optical measurements and acoustic reference values~\cite{hacker2021copolymer,grohl2023moving} to a manually created segmentation mask that delineates the different material regions of the real phantoms. \rev{Specifically, we simulate homogeneous phantoms with a density of 1000\,g/cm$^3$ and a sound speed of 1468\,m/s. In the coupling medium, we set a sound speed of 1489\,m/s and a density of 1000\,g/cm$^3$. We assume negligible acoustic attenuation.} We created multi-label segmentation masks using the medical imaging interaction toolkit (MITK)~\cite{nolden2013medical} based on delay-and-sum reconstructions of the measured time series data.

We simulated device-specific initial pressure distributions $p_0(x)$ and measurement data $g(t, y)$, based on the numerical phantoms. We first used a Monte Carlo model of light transport (MCX~\cite{fang2009monte}) to simulate the light fluence $\phi$ and calculated the expected initial pressure distribution using $p_0(x) = \Gamma \cdot \mu_a(x) \cdot \phi(x)$, assuming a constant Gr\"uneisen parameter $\Gamma$. We then use the k-space pseudospectral method implemented in k-Wave~\cite{treeby2010k} as the 3D acoustic forward model to generate measurement data $g(t, y)$ (see FIG~\ref{fig:digital_twin}).

\begin{figure}[h!tb]
    \centering
    \includegraphics[width=\linewidth]{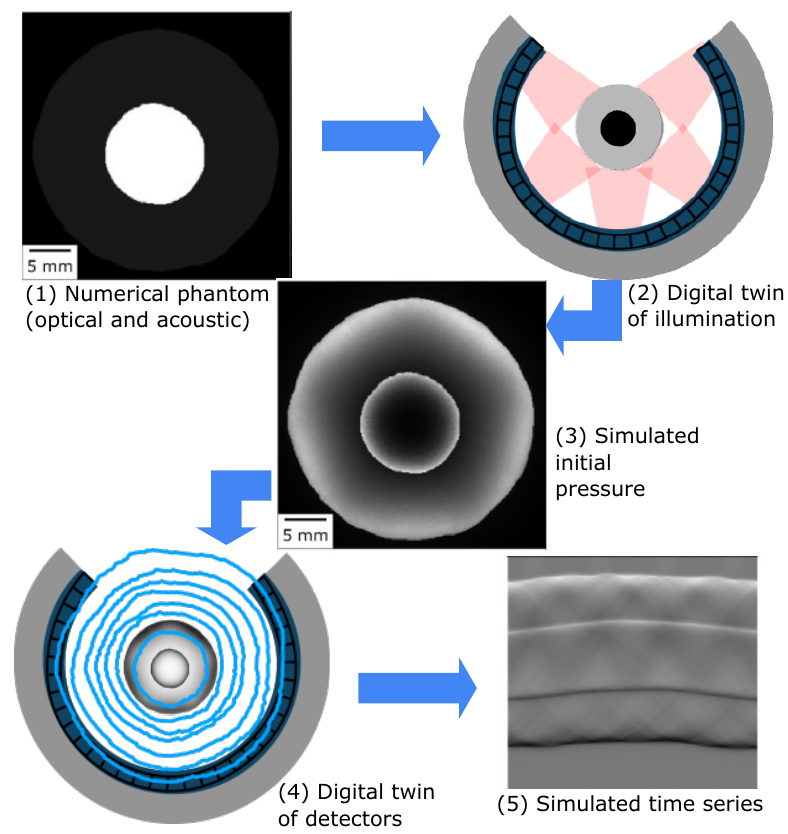}
    \caption{\textbf{Overview of the digital twin simulation pipeline.} (1) A numerical phantom with optical and acoustic properties matching a corresponding real phantom is created. (2) With a digital twin of the illumination geometry, the light fluence is computed using the Monte Carlo method. (3) The computed fluence is multiplied with the absorption coefficients to obtain the initial pressure distribution. (4) With a digital twin of the detection geometry, sound propagation is simulated using the k-space pseudospectral method, which leads to simulated measurements that correspond to experimental measurements.}
    \label{fig:digital_twin}
\end{figure}

We used a digital twin of the MSOT InVision-256TF that we implemented in the SIMPA toolkit~\cite{grohl2022simpa}. The computational model of the device was built using hardware geometry details provided by the vendor.} We implemented a custom device class in MCX and added a custom k-Wave Array definition of the detection elements of the MSOT InVision, which interpolates the toroidal surface of each transducer element with 707 points. We optimised the illumination and detector design parameters to best match our experiments based on N=15 calibration phantoms. \rev{We found good agreement between vendor specifications and the manually identified parameters and further found that a Gaussian illumination profile led to a good match between simulations and the experimental radiant exposure.} {The simulations were run using the open-source SIMPA~\cite{grohl2022simpa} toolkit, which essentially acts as an orchestration layer and provides adapters to both MCX and k-Wave.}\\

{We calibrated the proposed digital twin pipeline against experimental data using three components: a linear scaling of the simulated signal amplitudes, the introduction of a noise model based on an experimental measurement of the device noise, and the consideration of the impulse response of the imaging system.
We optimised the following linear parameters $a$, $b$, and $c$ based on N=15 calibration phantoms using a least squares optimisation scheme:
\[
g(t,y)_\text{exp} = a + (b \cdot g(t,y)_\text{sim}) \ast IRF(t, y) + c \cdot \text{noise}(t, y),
\]
where $g(t,y)_\text{exp}$ are the experimental measurements, $g(t,y)_\text{sim}$ are the outputs of the SIMPA simulation pipeline, a convolution with $IRF(t, y)$, the impulse response function of the imaging system that the manufacturer provided, and $\text{noise}(t, y)$ is a noise measurement taken in an empty water bath. We used the digital twin framework to compare the accuracy of this calibrated forward model to a na\"ive scaling of the SIMPA simulation, where the scaling factor was empirically determined to be 10.}

\subsection{Acoustic Inversion}

In the MSOT inVision system, the illuminated region is 
 concentrated around the plane $\Pi$ containing the detectors.
The detectors cover three fourths of a circle $S$ of radius $R=40.5$~mm lying in the plane. Our goal is to reconstruct the values of $p_0(x)$ for $x \in \Pi$ and restricted to the interior of circle $S$. 

Some of the known algorithms model this situation by assuming that
the region is infinitely thin, and use the 3D wave equation to describe 
the propagating wave. Others use the 2D model of wave propagation; this 
corresponds to the assumption that the region is extended in the direction
orthogonal to $\Pi$, and that $p_0(x)$ is invariant in that direction. \rev{In the following, we describe the reconstruction methods we use in this manuscript.}\\

%

\subsubsection{Circular FFT-based reconstruction} 

The Fourier series algorithm we investigate here experimentally is based on the method first proposed in~\cite{kunyansky2011reconstruction}, with additional recent correction based on the technique of~\cite{eller2020microlocally}. The method generates a theoretically exact inversion under the assumption that the direct problem is accurately modelled by a 2D wave equation, and the data is acquired by a full circle of detectors. In the situation when a certain subset of detectors is missing (as is the case with the MSOT scanner we use) the technique of~\cite{eller2020microlocally} allows us to significantly reduce the arising artefacts. {Additionally, the experimental system violates the ideal 2D assumption and hence an evaluation on experimental data is crucial to establish the algorithm's usefulness for application.}\\

For simplicity of presentation we scale the variables as follows: $\hat{t}=ct/R$, $\hat{x}=x/R,$ \ and introduce a function $u(\hat{t},\hat{x})=p(t,x).$ The function $u$ is a solution of the 2D wave equation in $\hat{t}$ and $\hat{x}$, with a unit speed of sound; also, $u(0,\hat{x})$ is supported in the unit circle, so that $p_{0}(x)=u(0,R\hat{x}).$ The algorithm consists of the following steps:\\

\begin{enumerate}[leftmargin=*,labelindent=0px]
\item Zero-pad data $g(t,y)$ in $t$ by a factor of 2 or more, by adding zeros
after the actually measured values.

\item Using the FFT, compute the Fourier transform of the zero-padded data
$g(ct/R,y(\theta))$ in $t$ and the Fourier series in $\theta$, where
$y(\theta)=R(\cos\theta,\sin\theta):$%
\[
\hat{g}_{k}(\rho)=\frac{c}{2\pi R}\int\limits_{\mathbb{R}}\left[
\int\limits_{0}^{2\pi}g(ct/R,\hat{y}(\theta))e^{-ik\theta}d\theta\right]
e^{it\rho}dt.
\]
The computed values are defined on a computational grid in $\rho\in
\lbrack-\rho_{\mathrm{Nyq}},\rho_{\mathrm{Nyq}}]_{,}$ $k\in\lbrack-N_d/2,N_d/2],$ where
$\rho_\mathrm{Nyq}$ is the Nyquist frequency of discretization in $\hat{t}$ and $N_d$ is the number of the detectors, assuming
that they fill the whole circle (we are only using positive
values of $\rho$). 

\item For each $k$ and each positive value of $\rho$ in the grid compute
coefficients $b_{k}(\rho)$ by the formula%
\[
b_{k}(\rho)=\frac{4i^{|k|}}{\rho H_{|k|}^{(1)}(\rho)}\hat{g}_{k}(\rho).
\]
where $H_m^{(0)}$ is the Hankels function of order $m$. 

\item For each value of $\rho$ in the grid, using the FFT, sum the Fourier
series thus obtaining function $\hat{B}(\rho,\varphi)$:
\[
\hat{B}(\rho,\varphi)=\sum\limits_{k=-N_d/2}^{N_d/2}b_{k}(\rho
)e^{ik\varphi}.
\]
Function $\hat{B}(\rho,\varphi)$ represents a theoretically exact polar grid representation to the Fourier transform $\hat{u}(0,\xi)$ of the function $u(0,\hat{x})$ we seek, assuming that $\xi=|\xi|(\cos\varphi,\sin\varphi)$, $\xi \in (0,\rho_{\max})$, $\varphi \in [0,2\pi)$.

\item Interpolate $\hat{B}(\rho,\varphi)$ from the polar grid to a Cartesian grid in
$\xi,$ producing an approximation to $\hat{u}(0,\xi).$ Here, we utilize bilinear interpolation.

\item Additional correction: the key observation of~\cite{eller2020microlocally} is that, if a segment of detectors is absent, roughly, a half
of the approximation to $\hat{u}(0,\xi)$ will be severely distorted, while the
other half will be reconstructed quite accurately. Since $u(0,\hat{x})$ is a
real function, its Fourier transform has the property $\hat{u}(0,\xi
)=\overline{\hat{u}(0,-\xi)}.$ The additional correction we deploy is to
replace the "bad" half of values $\hat{u}(0,\xi)$ by values of $\overline
{\hat{u}(0,-\xi)}$. 

\rev{Speaking crudely, this means that every point in the reconstruction image will receive 
its high spatial frequency information from the detectors it ``sees'' in the 180 degrees angular range. For points lying further away from the absent transducers this will mean fewer transducers; for points lying closer this will mean almost all of them. A more clear and rigorous explanation of this correction technique can be found in~\cite{eller2020microlocally}}.

\item Using a 2D FFT we compute $u(0,\hat{x})$ from $\hat{u}(0,\xi)$, and
reconstruct $p_{0}(x)=u(0,R\hat{x})$.\\
\end{enumerate}

{This FFT-based method, in its current implementation, is asymptotically fast, meaning that it requires $\mathcal{O}(n^2 \operatorname{log} n)$ floating point operations (flops) for an $(n \times  n)$ image, assuming that the data contains $\mathcal{O}(n^2)$ values. Other fast algorithms are only known for the cases of linear or planar arrangement of transducers, see for example \cite{jaeger2007fourier}. FFT reconstruction algorithms utilizing k-wave~\cite{treeby2010k} are not asymptotically fast in the sense that just one timestep of such an algorithm in 2D already requires $\mathcal{O}(n^2 \operatorname{log} n)$ flops. The present FFT-based method is available within the PATATO toolbox~\cite{else2024patato} and also on GitHub \url{https://github.com/leonidak/Fast_Hankel-based_solver}.}\\

\subsubsection{Delay-And-Sum} In delay-and-sum (DAS) reconstruction, the approximation $\tilde p_0(x)$ to $p_0(x)$ is calculated by summing all signals based on the time $\tau(x,y_j)$ it takes for the sound to travel from the spatial origin $x$ to the respective detector element $y_j$ on $S$~\cite{kirchner2018signed}:
\begin{equation*}
    \tilde p_0(x) = \sum\limits_{j=1}^{N_d} g(\tau(x, y_j), y_j),
\end{equation*}
where $y_j$ is the location of the $j$-th detector element on $S$ and $N_d$ is the number of these elements. We use the delay-and-sum implementation of the PATATO toolbox~\cite{else2024patato}.\\

\subsubsection{Filtered Backprojection} The filtered backprojection (FBP) implementation in PATATO applies the DAS reconstruction scheme to the time-derivative of the detected acoustic signals $g(t,y)$ after applying a low-pass filter $h$ that partially compensates for the frequency response of the imaging system:

\begin{equation*}
    p_0(x) = \sum\limits_{j=1}^{N_d} \frac{\partial}{\partial t} g(\tau(x, y_j), y_j) * h(t).
\end{equation*}

One should note that the Universal Backprojection as defined by Xu et al.~\cite{xu2005universal} contains a detector-dependent image weighting factor that was not considered in this implementation.\\

\subsubsection{Model-based reconstruction} 

Instead of using a simplified delay operator mapping $g(t,y_{j})$ into an
approximation to $p_{0}(x)$, model-based (MB) reconstruction techniques
explicitly incorporate a solution to the wave equation. We use an MB
algorithm provided in the PATATO toolbox~\cite{else2024patato}, which
implements the approach by Rosenthal et al.~\cite{rosenthal2010fast}.

The authors of {\cite{rosenthal2010fast}} approximate the solution to the 3D
wave equation~(\ref{E:3Dwave}) by 
implicitly {assuming that the initial
condition $p_{0}(x)$ is concentrated in an infinitely thin} region lying in
the plane $\Pi$. This leads to the following expression for the
measurements $g(t,y_{j})$, $j=1,...,N_{d}$:%
\begin{equation*}
g(t,y_{j})\thickapprox \frac{1}{c}\frac{\partial }{\partial t}%
\int\limits_{\mathbb{R}^{2}}\frac{p_{0}^{\mathrm{interp}}(x)\delta
(|y_{j}-x|-t)}{4\pi |y_{j}-x|}\,dx, 
\end{equation*}%
where $p_{0}^{\mathrm{interp}%
}(x)$ is a function of a 2D variable representing an approximation to $%
p_{0}(x)$ (with $x\in \mathbb{R}^{2}$). Function $p_{0}^{\mathrm{interp}}(x)$ is
defined by interpolating the values of $p_{0}(x_{k})$ given at the nodes $%
x_{k}$ of the discretization grid.{\ This leads to a system of linear
equations relating measurements }$g(t,y_{j})$ to the values of $%
p_{0}(x_{k})$ at the nodes, which is solved with any standard
approach, such as, for example, least squares. \\

\subsubsection{Time Reversal}

Our Time Reversal (TR) reconstruction is based on a two-dimensional
approximation to the forward problem. It is represented by equation (\ref{E:3Dwave}) with
$D$ equal to $2$. Such a model implicitly assumes that the initial pressure
$p_{0}(x)$ is invariant in the direction normal to plane $\Pi$, and the
detectors measure values of the pressure at the points sampling the whole circle
$S.$ By restricting the solution $p(t,x)$ of the forward problem to the
interior $B$ of circle $S$, one can see that $p$ is a solution of the
initial/boundary value problem in the time-space cylinder 
$Z=(0,T)\times B$ with initial conditions $p(0,x)=p_{0},$ $p_{t}(0,x)=0$ in $B$, and boundary values
$p(t,y)=g(t,y)$ at all $(t,y)\in(0,T)\times S_{.}$ One also observes that, in
practice, for a sufficiently large $T$, values of $p(T,x)$ and $p_{t}(T,x)$ are quite small and can be approximated by 0.
The idea of the TR \cite{ak07,Kun-series,Hristova2008} is to solve
the wave equation backwards in time, in the time-space cylinder $(t,x)\in
(0,T)\times B,$ with the boundary conditions $g$ and zero initial conditions
at $t=T.$ In other words, one would like to find the solution $u(t,x)$ to the
following problem:%
\begin{equation}%
\begin{cases}
u_{tt}=c^{2}\Delta u,\quad(t,x)\in(0,T)\times B,\\
u(T,x)=0,\quad u_{t}(T,x)=0,\\
u(t,y)=g(t,y),\quad(t,y)\in(0,T),
\end{cases}
\label{E:goodTR}%
\end{equation}
with the approximation $\tilde{p}_{0}$ to $p_{0}$ computed as
\[
\tilde{p}_{0}(x)=u(0,x),\quad x\in B.
\]
Ideally, if $g(t,y)$ is known exactly on $(0,T)\times S$, in the limit
$T\rightarrow\infty,$ the approximation $\tilde{p}_{0}$ becomes exact, i.e.
$p_{0}(x)=u(0,x).$ When a part of the circle $S$ does not contain the detectors,
the corresponding values are replaced by $0$, and $\tilde{p}_{0}$ represents a
crude approximation to $\tilde{p}_{0}.$ Using the solution of problem
(\ref{E:goodTR}) one defines an operator $A^{\triangleleft}$ mapping
boundary values $g$ into $\tilde{p}_{0}$:%
\[
\mathcal{A}^{\vartriangleleft}:g\rightarrow\tilde{p}_{0}.
\]
So, our TR technique consists of computing $\tilde{p}_{0}=\mathcal{A}%
^{\vartriangleleft}g$, by solving numerically problem (\ref{E:goodTR}). For
the present study we utilize the implementation of TR described in the
Appendix of \cite{arridge2016adjoint} that, in turn, is based on the use of the 2D version of the k-Wave
toolbox~\cite{treeby2010k}.



\subsubsection{Iterative Time Reversal} The iterative time reversal (ITTR) method \cite{stefanov2009thermoacoustic} uses TR on each iteration step, to refine
the current approximation to $p_{0}$. The next approximation $\tilde{p}_{0}^{(k+1)}(x)$
is defined through the previous approximation $\tilde{p}_{0}^{(k)}(x)$ by
the formula%
\begin{equation}
\tilde{p}_{0}^{(k+1)}=\tilde{p}_{0}^{(k)}+\mathcal{A}^\triangleleft \left( g-\mathcal{A}
\tilde{p}_{0}^{(k)}\right) ,\quad k=0,1,2,3,...
\label{E:iterative}
\end{equation}%
with $\tilde{p}_{0}^{(0)}=0$ and where $\mathcal{A}$ is the operator that solves (\ref{E:3Dwave}).


In this case $\tilde{p}_{0}^{(1)}=\mathcal{A}%
^{\triangleleft }g$ is just the approximation obtained by the TR algorithm. As in TR, the application of the operator $\mathcal{A}^{\triangleleft }$ is equivalent to solving problem (\ref{E:goodTR}) with the boundary condition $g-\mathcal{A}\tilde{p}_{0}^{(k)}$. We utilize the k-Wave toolbox to implement each step of this algorithm.

\subsection{Image Quality Assessment Measures}

We use five full-reference image quality assessment (IQA) measures and one no-reference measure to compare the reconstruction performance of the different reconstruction methods. We chose the IQA measures to capture different properties and relationships between the images. Each is described in the following\rev{, where $N$ is the number of pixels in the image}:\\

\subsubsection{\rev{The Pearson Correlation Coefficient} ($R$)} $R$ quantifies the strength and direction of the linear relationship between two variables. It ranges from -1 to 1, where 1 indicates a perfect positive linear relationship, -1 indicates a perfect negative linear relationship, and 0 indicates no linear relationship. We use the coefficient to quantify the strength of the linear correlation between the reconstructed signal $\tilde p _0$ and the simulated initial pressure distribution $p_0$~\cite{cohen2009pearson}.

\begin{equation*}
R = \frac{\sum_{i=1}^N \left(\tilde{p}_{0,i} - \overline{\tilde{p}_0}\right) \left(p_{0,i} - \overline{p_0}\right)}{\sqrt{\sum_{i=1}^N \left(\tilde{p}_{0,i} - \overline{\tilde{p}_0}\right)^2} \sqrt{\sum_{i=1}^N \left(p_{0,i} - \overline{p_0}\right)^2}}
\end{equation*}

\subsubsection{\rev{Mean Absolute Error} (MAE)} The MAE quantifies the average magnitude of errors between predicted and actual values, without considering their direction. It is calculated by taking the average of the absolute differences between 
$\tilde p _0(x)$ and $p_0(x)$, where $x$ represents the pixel position.

\begin{equation*}
    \text{MAE} = \frac{1}{N} \sum_{x=1}^N \left| \tilde{p}_{0}(x) - p_{0}(x) \right|
\end{equation*}

\subsubsection{\rev{Structural Similarity Index Measure} (SSIM)} The Structural Similarity Index Measure (SSIM) evaluates the similarity between two images by jointly comparing their structural information, luminance, and contrast. The SSIM between $\tilde p _0$ and $p_0$ can be calculated using the mean, variance, and covariance of the images~\cite{wang2004image}.

\begin{equation*}
    \text{SSIM} = \frac{(2 \cdot \overline{\tilde{p}_0} \cdot \overline{p_0} + C_1)(2 \sigma_{\tilde{p}_0 p_0} + C_2)}{(\overline{\tilde{p}_0}^2 + \overline{p_0}^2 + C_1)(\sigma_{\tilde{p}_0}^2 + \sigma_{p_0}^2 + C_2)},
\end{equation*}
where $\overline{x}$ denotes the mean, $\sigma^2$ denotes the variance and $\sigma_{\tilde{p}_0 p_0}$ the covariance of $\tilde{p}_0$ and $p_0$. $C_1$ and $C_2$ are stabilising constants. \rev{We use the TorchMetrics default parameters of C1 = 0.01 and C2 = 0.03~\cite{torchmetrics2022}}.

\subsubsection{\rev{Jensen-Shannon Divergence} (JSD)} The JSD quantifies the similarity between two probability distributions. It is a symmetric and smoothed version of the Kullback-Leibler divergence and provides a bounded value between 0 (when the distributions are identical) and 1 (when the distributions are maximally different)~\cite{menendez1997jensen}. We perform $z$-score normalization for all pixel values in $\tilde p _0$ and $p_0$ and create histograms P and Q with N=100 bins ranging from $-3\sigma$ to $3\sigma$ to extract the probability distributions. We use a Python implementation of the Jensen-Shannon distance, available in the Scipy (v1.10.1) package~\cite{virtanen2020scipy}.

\[
\text{JSD}(P \parallel Q) = \frac{1}{2} \text{KL}(P \parallel M) + \frac{1}{2} \text{KL}(Q \parallel M)
\]
where
\[
M = \frac{1}{2}(P + Q)
\]
and
\[
\text{KL}(P \parallel M) = \sum_{i=1}^N P(i) \log \frac{P(i)}{M(i)}
\]
\[
\text{KL}(Q \parallel M) = \sum_{i=1}^N Q(i) \log \frac{Q(i)}{M(i)}
\]

\subsubsection{\rev{Haar Wavelet-based Perceptual Similarity Index} (HaarPSI)} HaarPSI assesses the visual similarity between two images \rev{in the interval [0, 1]}. It leverages the Haar wavelet transform to capture the local phase coherence and contrast of the images. HaarPSI between the reconstructed signal \(\tilde p _0 \) and the simulated initial pressure distribution \( p_0 \) is calculated based on the responses of the Haar wavelet transform applied to both images~\cite{reisenhofer2018haar}. We use the Python implementation of the method available at \url{https://github.com/rgcda/haarpsi}.

\begin{equation*}
    \text{HaarPSI} = l_{\alpha}^{-1} \left( \frac{\sum_{x} \sum_{k=1}^{2} \text{HS}_{p_0, \tilde{p}_0}^{(k)}[x] \cdot W_{p_0, \tilde{p}_0}^{(k)}[x]}{\sum_{x} \sum_{k=1}^{2} W_{p_0, \tilde{p}_0}^{(k)}[x]} \right)^2,
\end{equation*}

where $\alpha > 0$ is a free parameter (here $\alpha = 4.2$), $l_{\alpha}(x) = 1 / (1 + e^{-\alpha x})$, $W$ is a weight map derived from the response of a single low-frequency Haar wavelet filter $k$, and HS is the local similarity measure which is based on the first two stages of a two-dimensional discrete Haar wavelet transform.\\ 

\subsubsection{\rev{Full Width at Half Maximum} (FWHM)} The FWHM is a no-reference IQA measure and quantifies the width of signal peaks. It is defined as the distance between the points on a signal curve at which the signal amplitude falls to half of its maximum value~\cite{castleman1996digital}. We use it as a measure to quantify the sharpness of the reconstruction by computing the FWHM of the absolute gradient along a manually defined line profile through the object boundaries in $\tilde p _0$. We compute the FWHM using the following steps:

\begin{enumerate}
    \item Extract a line profile through the image that includes both background and target boundaries.
    \item Compute the absolute value of the gradient of the line profile.
    \item Identify the peaks in the absolute gradient.
    \item For each peak, locate the first positions to the left and right where the absolute gradient drops to half of the peak value.
    \item Compute the width by subtracting the left position from the right position.
\end{enumerate}

\subsection{Computational Footprint Estimation}

We evaluated the computational footprint, including both time and memory usage, for the specific algorithm implementations. To measure execution time, we used timing functions in MATLAB and Python. For the MATLAB script, we recorded the peak memory consumption of the Graphics Processing Unit (GPU), as reported by k-Wave. In Python, we used the \textit{tracemalloc} module around the execution block to track the peak memory consumption of the Central Processing Unit (CPU), and the \textit{nvidia-smi} tool to monitor peak GPU memory consumption. Each algorithm was tested by reconstructing a single data point three times. We computed the total script execution time of each run, which includes any data loading operations as well as the initialisation of any Python or MATLAB modules. We run these experiments on an Intel(R) Core(TM) i9-10900KF CPU @ 3.70GHz CPU with 64 GB RAM and an NVIDIA RTX3090 graphics card.

\section{Results}
\label{sec:results}

\subsection{With the digital twin pipeline, the accuracy of the forward model can be evaluated and compared.}

\begin{figure}[h!tb]
    \centering
    \includegraphics[width=\linewidth]{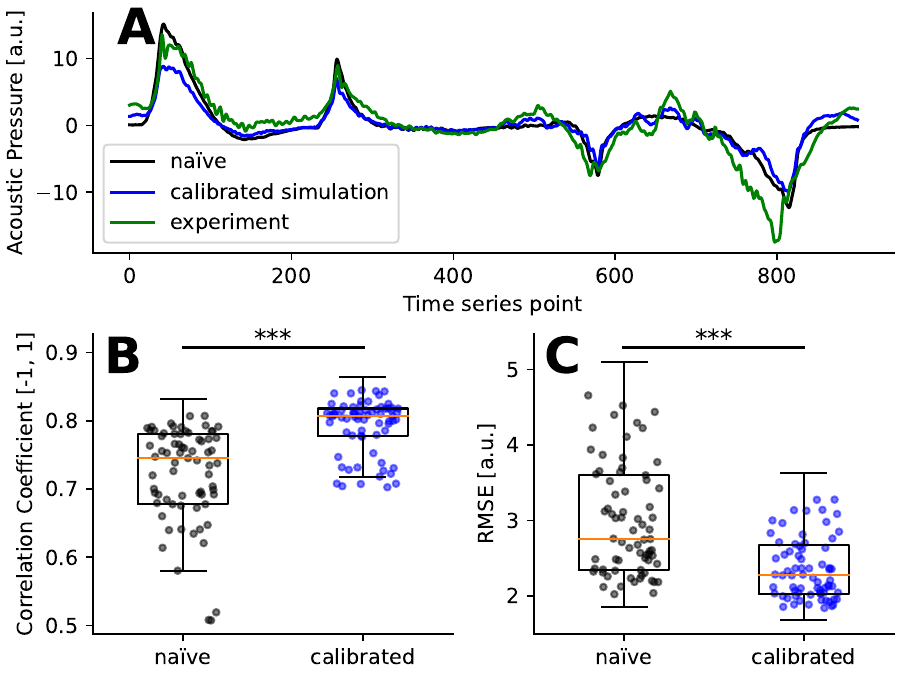}
    \caption{\textbf{Our calibration scheme achieves higher agreement with the experiment compared to na\"ive scaling of the simulation.} We first show a qualitative comparison of the measurement data compared to na\"ive and calibrated simulations of a representative measurement of a single detector of a randomly chosen test phantom (\textbf{A}). We then use the corresponding measurement pairs between simulation and experiment to show that we can evaluate the accuracy of the forward model, for example, in terms of the Pearson Correlation Coefficient (\textbf{B}) or the root mean squared error (\textbf{C}). Both of these evaluations show that a careful calibration scheme that includes noise modelling and modelling of the impulse response significantly outperforms a na\"ive scaling of the k-Wave simulations. *** indicates p<0.0001 using a Wilcoxon signed-rank test.}
    \label{fig:res:pipeline}
\end{figure}

{To investigate the accuracy of our forward model, we compared the simulated measurement data $g(t,y)_\text{sim}$ with the experimental measurements $g(t,y)_\text{exp}$ in the time domain. Visual inspection of the data shows generally good agreement between the two (FIG~\ref{fig:res:pipeline} A), where one can observe peaks and valleys of the pressure waves in the same locations and with similar amplitudes. Also the noise patterns have a high degree of similarity. }

{In addition to qualitative evaluation, we can use the $g(t,y)_\text{sim}$ and $g(t,y)_\text{exp}$ data pairs to quantify the difference and to compare the goodness of fit of different forward models. To this end, we compared two calibration approaches: (1) a na\"ive scaling of the SIMPA simulated $g(t,y)_\text{sim}$, empirically determined as $g(t,y)_\text{na\"ive} = g(t,y)_\text{sim} / 10$ and (2) an optimised calibration: $g(t,y)_\text{cal} = 4.5 + (0.068 \cdot g(t,y)_\text{sim}) \ast IRF(t, y) + 0.89 \cdot \text{noise}(t, y)$. With this calibration, we could improve the Pearson correlation coefficient from 0.75 to 0.81 (FIG~\ref{fig:res:pipeline} B) and reduce the root mean squared error from 2.75 to 2.27\,a.u. (FIG~\ref{fig:res:pipeline} C).}

\subsection{Qualitative and quantitative performance comparison reveals similar reconstruction quality of the FFT-based method}

\begin{figure}[!ht]
    \centering
    \includegraphics[width=\columnwidth]{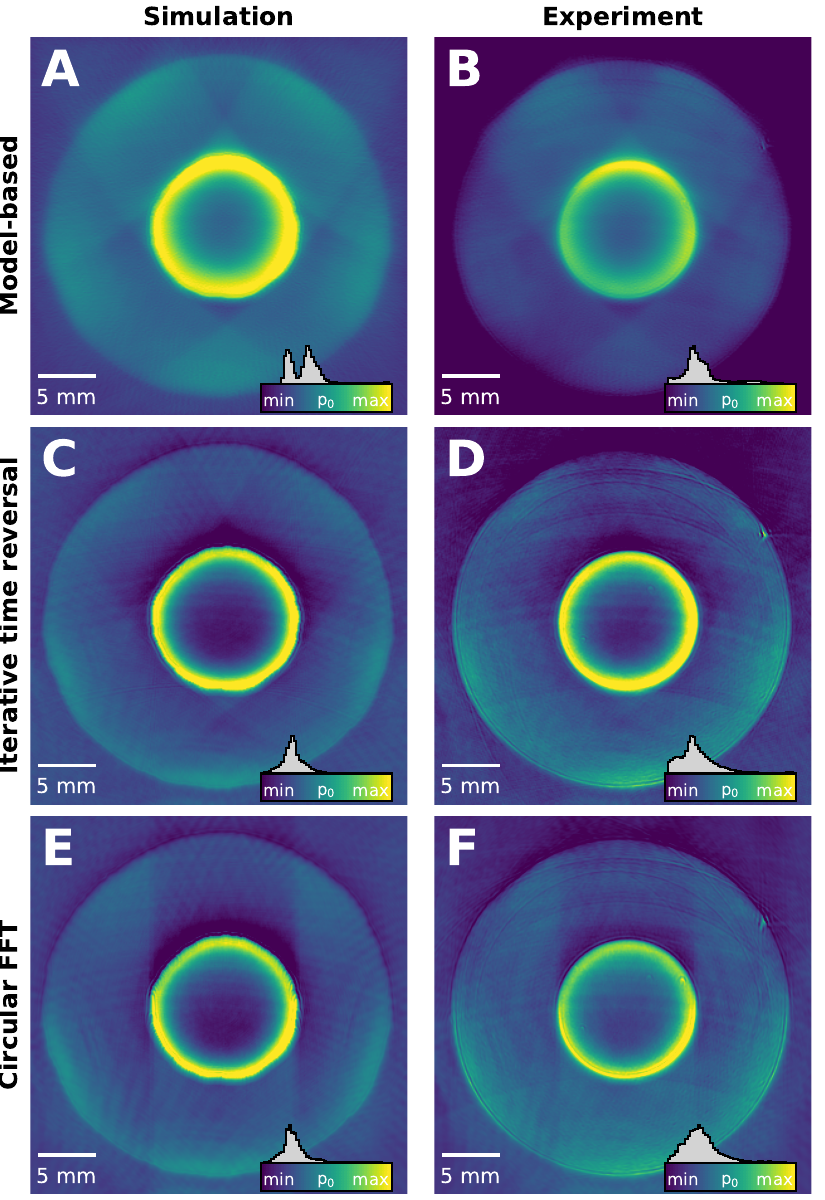}
    \caption{\textbf{Qualitative visualisation of an example phantom with a circular imaging target for simulated and experimentally acquired data.} The figure shows reconstruction results for three algorithms: model-based reconstruction (A, B), iterative time reversal (C, D), and the FFT-based method (E, F). The reconstructions are shown for simulated data (A, C, E) and experimentally acquired data (B, D, F). Each panel shows a value histogram with the reconstructed values relative to the simulated initial pressure distribution.}
    \label{fig:ex_2}
\end{figure}

We first qualitatively show the reconstruction results of a representative test phantom (cf. FIG.~\ref{fig:ex_2}). We show the results of the idealised simulations (panels A, C, E) side-by-side with the experiments (panels B, D, F). The results show that reconstruction artefacts are more pronounced in the simulations across all methods (see esp. Fig~\ref{fig:ex_2} A, C, E) and that there is a significant level of noise in the experimental images. Furthermore, the qualitative comparison suggests that the circular FFT-based method produces results that are similar to the computationally expensive iterative time reversal method.

For quantitative comparison, we compute full-reference IQA measures that relate the reconstructed image to the \textit{in silico} initial pressure distribution which was obtained from numerical simulation of the digital twins. In addition, we compute the FWHM as a no-reference IQA measure that is indicative of the sharpness of the reconstructions. Our results show that {time reversal and iterative time reversal} perform nearly equivalently and best across nearly all measures (cf. Table~\ref{tab:sim}). The only notable exceptions {are HaarPSI, which favours model-based reconstruction and filtered backprojection, and} the FWHM, which indicates that the FFT-based method can reconstruct the sharpest object boundaries - an observation that is corroborated by the qualitative results.

\begin{table}[!ht]
	\centering
	\begin{tabular}{lllllll}
		Experiment & \textbf{DAS} & \textbf{FBP} & \textbf{MB} & \textbf{TR} & \textbf{ITTR} & \textbf{FFT}\\ \hline
		\\
		\textbf{R} $\uparrow$ & \cellcolor{orange!75} 0.24 & \cellcolor{orange!25} 0.57 & 
 \cellcolor{orange!25} 0.57 & \cellcolor{gree!50} 0.76 & \cellcolor{gree!75} \textbf{0.77} & \cellcolor{gree!25}0.61\\ 
		\textbf{MAE} $\downarrow$ & \cellcolor{orange!75} 199.84 & \cellcolor{orange!50} 163.18 & \cellcolor{orange!25} 156.39 & \cellcolor{gree!75} \textbf{109.43} & \cellcolor{gree!50} 110.91 & \cellcolor{gree!25} 140.17\\ 
		\textbf{SSIM} $\uparrow$ & \cellcolor{orange!25} 0.72 & \cellcolor{orange!25} 0.72 & \cellcolor{orange!75} 0.69 & \cellcolor{gree!50} 0.77 & \cellcolor{gree!75} \textbf{0.78} & \cellcolor{gree!25} 0.75\\ 
		\textbf{JSD} $\downarrow$ & \cellcolor{orange!75} 0.48 & \cellcolor{orange!25} 0.43 & \cellcolor{orange!50}0.44 & \cellcolor{gree!75}\textbf{0.36} & \cellcolor{gree!50}0.37 & \cellcolor{gree!25}0.41\\ 
		\textbf{HaarPSI} $\uparrow$ & \cellcolor{orange!25} 0.33 & \cellcolor{gree!75} \textbf{0.50} & \cellcolor{gree!50} \textbf{0.46} & \cellcolor{gree!25}0.37 & \cellcolor{orange!75} 0.31 & \cellcolor{orange!50} 0.32\\ 
		\textbf{FWHM} $\downarrow$ & \cellcolor{orange!75} 8.20 & \cellcolor{orange!50} 7.35 & \cellcolor{orange!25} 6.42 & \cellcolor{gree!50} 4.79 & \cellcolor{gree!25}4.92 & \cellcolor{gree!75}\textbf{4.55}\\ 
	\end{tabular}
\caption{\label{tab:sim} \textbf{Quantitative IQA reveals algorithm performance differences on experimental data.} The tables show the quantitative results for all measure/algorithm combinations. $\uparrow$ denotes that higher values are better and $\downarrow$ denotes that lower values are better. Bold values denote the best-performing algorithm. The relative performance of the algorithms is also denoted as the cell colour where the top three algorithms are labelled in shades of green and the bottom three are labelled in shades of orange.}\end{table}

Notably, the FFT-based method only performs slightly worse compared to the TR and ITTR methods and significantly outperforms model-based reconstruction and filtered backprojection. Delay-and-sum reconstruction without signal processing performs significantly worse than all other methods. Our results show that on experimental data, the FFT-based algorithm achieves performance measures that are {worse compared to the time reversal algorithms, but better than backprojection and filtered backprojection algorithms}.

\subsection{Reconstruction algorithms show differences when moving from a full-view to a limited-view setting.}

{We evaluate the change in performance of the MB, ITTR, and FFT method when moving from a full-view scenario (360$^\circ$ angular coverage) to a limited view setting (270$^\circ$ angular coverage, as encountered in the MSOT InVision scanner). We report the average results on 15 simulated phantoms at 800\,nm randomly chosen from the calibration and testing data.}

\begin{figure}[!ht]
    \centering
    \includegraphics[width=\columnwidth]{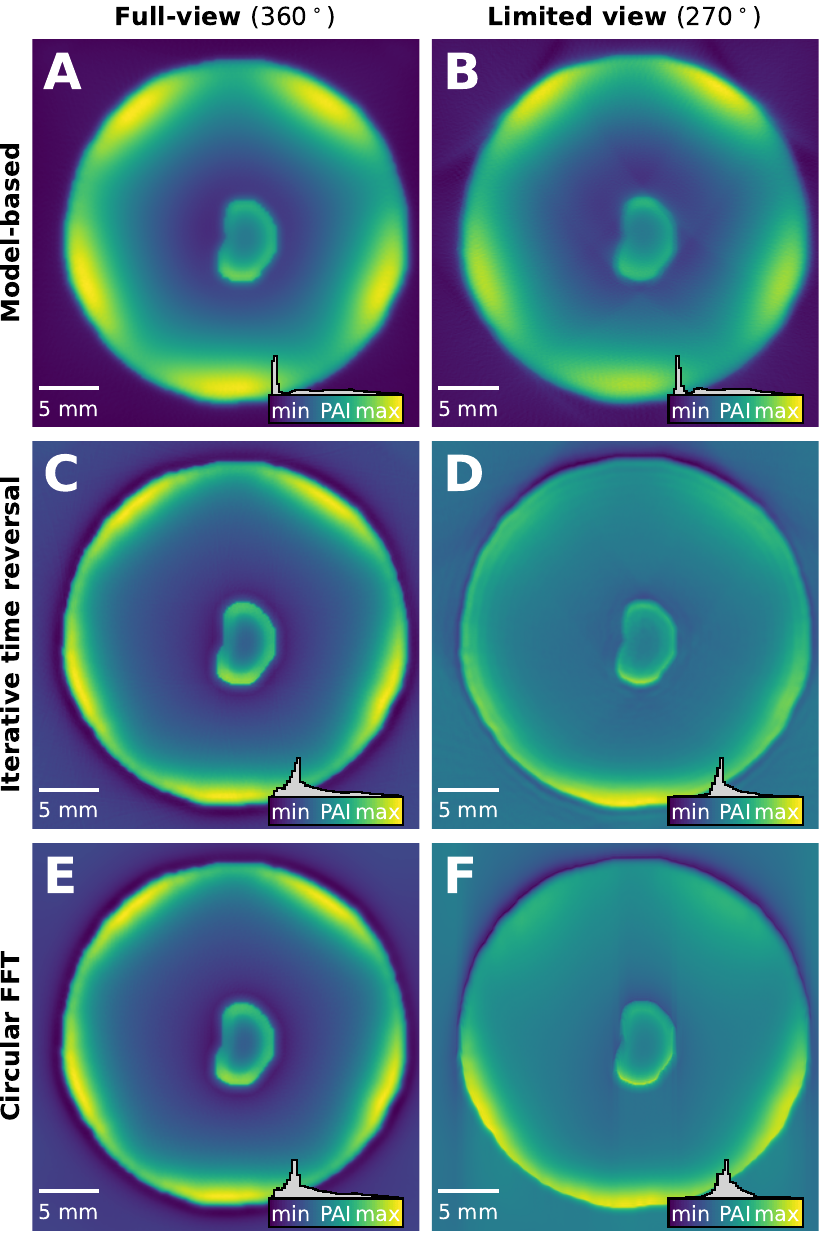}
    \caption{\textbf{Comparisons of full view versus limited view reconstructions reveal the emergence of reconstruction artefacts in the limited view case.} The figures show a representative example phantom simulation in a full-view setting (\textbf{A, C, E}), and a limited view setting (\textbf{B, D, F}). We show the results of the model-based reconstruction \textbf{(A, B)}, iterative time reversal \textbf{(C, D)}, and the FFT-based method \textbf{(E, F)}.}
    \label{fig:limited_view}
\end{figure}

\begin{table}[!ht]
	\centering
	\begin{tabular}{l|ll|ll|ll|}
		\multicolumn{1}{c}{} & \multicolumn{2}{c}{\textbf{MB}} & \multicolumn{2}{c}{\textbf{ITTR}}  & \multicolumn{2}{c}{\textbf{FFT}} \\ 
        & 360$^\circ$ & $\Delta$ \% & 360$^\circ$ & $\Delta$ \% & 360$^\circ$ & $\Delta$ \%\\

		\multicolumn{7}{c}{}\\
		\textbf{R} $\uparrow$ & 0.89 & +1.6\% & 0.94 & -6.6\% & 0.93 & -10.6\%\\ 
		\textbf{MAE} $\downarrow$ & 43.0 & -3.4\% & 88.38 & +8.2\% & 91.88 & +10.7\%\\ 
		\textbf{SSIM} $\uparrow$ & 0.87 & +1.3\% & 0.86 & -1.4\% & 0.85 & -2.4\%\\ 
		\textbf{JSD} $\downarrow$ & 0.24 & -2.3\% & 0.29 & +14.6\% & 0.28 & +11.0\%\\ 
		\textbf{HaarPSI} $\uparrow$ & 0.57 & +10.4\% & 0.69 & -12.3\% & 0.65 & -13.6\%\\ 
	\end{tabular}
\caption{\label{tab:limitedview}  \textbf{Reconstruction algorithms show performance differences in limited view settings.} We compare the performance measures of the model-based reconstruction (MB) with iterative time reversal (ITTR) and the Fourier transform-based method (FFT). For each, we show the measure scores for the full-view case (360$^\circ$) and the relative performance change in per cent ($\Delta \%$) when switching to a limited view setting with 270$^\circ$ coverage. $\uparrow$ denotes that higher values and positive changes are better and $\downarrow$ denotes that lower values and negative changes are better.}\end{table}
{We can make two immediate observations when visually inspecting the reconstruction results: (1) the MB method does not seem to notably change visually (cf. FIG.~\ref{fig:limited_view} A,B), whereas (2) the ITTR and FFT methods seem to deteriorate in a similar fashion (cf. FIG.~\ref{fig:limited_view} C-F). These findings are corroborated by the quantitative results (Table~\ref{tab:limitedview}). For the MB method, R, \rev{MAE}, SSIM, and JSD only slightly change and HaarPSI even shows improvements. For ITTR and FFT, on the other hand, all measures, except for SSIM, show a deterioration of the reconstruction performance in the order of 10\% (varying from 7\% to 15\%). One should note that the PATATO implementation of the MB method was exclusively tested against limited-view data from the MSOT InVision 256-TF, which might have introduced a bias that could constitute a reason why it might not deteriorate similarly compared to the others.}

{A second important observation is that the SSIM measure does not seem to reflect the changes to the same extent as the other measures. Some reconstruction artefacts around the central absorber in the image are visible for the MB method, but only the JSD measure predicts a deterioration of the image quality in all three cases. This indicates that there are stark differences in the type of data and changes that the different measures are sensitive to.}


\subsection{Computation analysis confirms high computational efficiency of the FFT-based method}

The full-pipeline run times for DAS, FBP, and TR were in the order of 2-3 seconds, whereas the runtimes of MB and ITTR were in the order of one minute per image. The FFT-based algorithm took an average of 1.4 seconds per image. A noteworthy observation is that the FFT-based algorithm only required a peak memory allocation of 200MB, whereas all other methods required 1.5GB and more. That being said, it must be noted that the algorithms' run times and memory requirements are heavily implementation-dependent. Our versions of DAS, FBP, and MB are implemented in Python in the research toolkit PATATO~\cite{else2024patato} and use Jax to run on the GPU. TR and ITTR are implemented in MATLAB and use the k-Wave toolbox which uses CUDA-compiled binaries. The FFT-based algorithm is implemented in Python and runs on the CPU. It has been demonstrated that through meticulous optimisation and e.g. the precomputation of delays or model matrices, the computational performance can be heavily reduced and single-stage reconstruction schemes can typically be applied in real time~\cite{yuan2013real}. The values of the Hankel functions $H_{|k|}(\rho)$ at the nodes of the equispaced Cartesian grid in $\rho\in\lbrack-\rho_{\mathrm{Nyq}},\rho_{\mathrm{Nyq}}]_{,}$ $k\in\lbrack-N_d/2,N_d/2])$ within the FFT-based method can be precomputed, which could significantly speed up the processing time. 

\section{Discussion}
\label{sec:discussion}

{Based on our results,} we believe that evaluating image reconstruction algorithms with full-reference IQA measures based on digital twins of the test objects and imaging system is an objective way to ensure comparability of reconstruction algorithms that tackle the acoustic inverse problem in photoacoustic imaging. No-reference IQA measures {do not directly indicate the accuracy of the initial pressure reconstruction}, though they can be a valuable complement to full-reference IQA measures. Since the ground truth $p_0$ is not known \textit{in vivo}, a high-confidence reference $p_0$ simulation for a real-world test object is therefore an ideal middle ground that enables objective comparison of acoustic inversion schemes using full-reference IQA measures and experimentally acquired data.

The proposed phantom materials by Hacker et al.~\cite{hacker2021copolymer} are accurately characterisable with a DIS measurement system and it is straightforward to manufacture complex piecewise-constant test objects. We can build digital twins of the phantoms and run simulations using the proposed digital twin simulation pipeline within SIMPA. {By comparing the simulated versus experimentally determined measurement data, we can quantitatively evaluate the fidelity of the implemented forward model and then use intermediate simulation outputs, such as the initial pressure distribution, as high-fidelity data for full-reference IQA measures. \rev{Other non-linear calibration functions are also valid choices, but in our experiments none of the tested functions yielded improvements significant enough to favour over a simple linear model. We believe this is because the impulse response and a measured noise profile of the imaging system have the biggest influence on the simulation fidelity - in conjunction with scaling the simulation to match the experimental amplitudes.}} \rev{Our approach is generalizable to any photoacoustic system, by manufacturing and measuring custom phantoms and modelling the PAI device in question in, e.g., MCX and k-Wave.} \rev{We further believe that our proposed approach can generalise to any type of complexity in the phantoms. While more complex phantoms could lead to more representative performance analyses, the added complexity in manufacturing and segmentation might undermine the general feasibility of the approach.}

We demonstrate the usefulness of the ability to apply full-reference IQA measures by comparing an FFT-based image reconstruction algorithm that was conceived to work with circular detection geometries to several state-of-the-art image reconstruction techniques. With the proposed framework we can show that the method exhibits similar performance as the iterative time reversal algorithm while being considerably faster. We further show that {we can use the digital twin framework to evaluate the performance changes of the reconstruction schemes when moving from a full-view to a limited-view setting. We find that the model-based reconstruction algorithm behaves drastically differently compared to iterative time reversal and the FFT-based method. At the same time, we also find that SSIM does not manage to pick up on the performance changes to the same extents that the other IQA measures under investigation did. This might indicate that SSIM is not a good fit to evaluate PA image reconstruction algorithms.}

One has to keep in mind that the simulated $p_0$ is only a reference and not an exact ground truth. The exact quantitative results of the IQA measures depend on the correctness of the forward model, {which hinges mainly on the approximating assumptions done in the mathematical models, but also on the representation of hardware constraints such as noise and the impulse response function. One can see this \textit{simulation gap} quantitatively when comparing the simulated and experimental measurement data. We show that careful calibration of the simulations with the reconstructions can correct some} modelling errors, but systematic non-linear changes could not be captured. Furthermore, the piecewise-constant nature of the proposed test objects is not sufficient to capture the complexity of in vivo tissue and sophisticated fabrication strategies, such as 3D printing approaches, would be required to achieve the spatial heterogeneity needed. The complementary information of the FWHM measure {and the lack of descriptive power of the SSIM measure in this context} also shows that the suite of IQA measures should probably be tailored to the target application. Ideally, the PAI community should identify common use cases for PAI and define a ranked list of the most useful IQA measures that capture all relevant aspects.

\section{Conclusion}
\label{sec:conclusion}

We demonstrate that digital twins can enable full-reference quality assessment of reconstructed photoacoustic images by simulating numerical equivalents of tissue-mimicking phantoms and the imaging system. With these, we propose a unique evaluation framework that {can compare algorithm performance both within simulated and experimental setting and we quantitatively evaluate different image reconstruction algorithms.}
We believe the approach can provide a robust and objective method for evaluating acoustic inversion schemes, enhancing reproducibility and accuracy in photoacoustic imaging. Furthermore, our study highlights {successful application of an FFT-based reconstruction method that performs comparably to state-of-the-art approaches with high computational efficiency (up to $100$-times faster) on the experimental data}. This is especially promising for developing model-based learned reconstruction methods, which necessitate using highly efficient models.

\begin{acknowledgments}
J.G. is funded by the Deutsche Forschungsgemeinschaft (DFG, German Research Foundation) under projects GR 5824/1 and GR 5824/2. L.K. is partially supported by the NSF, through the award NSF/DMS-2405348.
T.R.E. is funded by Cancer Research UK (A29580).
A.H. acknowledges support from the Research Council of Finland with the Flagship of Advanced Mathematics for Sensing Imaging and Modelling proj. 359186. Centre of Excellence of Inverse Modelling and Imaging proj. 353093, and the Academy Research Fellow proj. 338408. 
J.P. acknowledges support from the European Research Council (ERC) under
the European Union’s Horizon 2020 research and innovation programme (grant agreement No 101001417 - QUANTOM), the Research Council of Finland (Centre of Excellence in Inverse Modelling and Imaging grant 353086 and Flagship of Advanced Mathematics
for Sensing Imaging and Modelling grant 358944) and the Finnish Ministry of Education and Culture’s Pilot for Doctoral Programmes (Pilot project Mathematics of Sensing, Imaging and Modelling).
BTC acknowledges support from the 
Engineering and Physical Sciences Research Council, UK (EPSRC), grants EP/W029324/1, EP/T014369/1.
The authors would like to thank the Isaac Newton Institute for Mathematical Sciences, Cambridge, 
for support during the `Rich and Nonlinear Tomography' programme, supported by EPSRC grant EP/R014604/1.
\end{acknowledgments}

\section*{Author declaration}

The authors have no conflict of interest to declare.

\section*{Data availablility}

\rev{All data and code to reproduce this work are publicly available on Zenodo: \url{https://doi.org/10.5281/zenodo.15388429}}.

\bibliography{bibliography.bib}

\end{document}